# N-body Systems with Linear Wind Resistance: Analytic Solutions

Joseph West*

*Department of Chemistry and Physics, Indiana State University*

(Dated January 6, 2026)

**Abstract**

The attractive and repulsive linear Hookean form of gravity is known to allow for analytic solutions to N-body systems for arbitrary masses and initial conditions. This linear system is very well suited for use in the advanced undergraduate classroom with the position of every particle $\mathbf{r}_j(t)$ known analytically in terms of only sine, cosine, and exponential functions. Here it is shown that simple analytic solutions are also possible when linear wind resistance is included, provided that the force due to wind resistance force on each particle is proportional to the mass of that particle. Surprising analytic expressions for the behavior of the system if only a single particle in the system is subject to a linear wind resistance are also presented.

## I. Equations of Motion

The equations of motion of a set of masses $\{m_j\}$ subject to a linear Hookean form of gravity and linear wind resistance are

$$\ddot{\mathbf{r}}_j \equiv -JM\mathbf{r}_j - b_j\mathbf{v}_j + JM\mathbf{R} \qquad (1)$$

where $\mathbf{r}_j$ is the position of the particle, J is a constant with units of N/kg$^2$m can be positive (attractive) or negative (repulsive), M is the total mass, $b_j$ is a constant,[2] $\mathbf{v}_j$ is the velocity of particle j, and $\mathbf{R}$ is the position of the center of mass.[1] It is assumed that the wind resistance is proportional to the mass of the particle and the equations of motion become

$$M\ddot{\mathbf{R}} = -b_j\mathbf{v}_j \equiv -\sum_j 2\Gamma m_j \dot{\mathbf{r}}_j = -2\Gamma M\dot{\mathbf{R}} \equiv -2\Gamma M\mathbf{V} \qquad (2)$$

$$\ddot{\mathbf{r}}_{aj} = -w^2 \mathbf{r}_{aj} - 2\Gamma\dot{\mathbf{r}}_{aj} + w^2\mathbf{R} \qquad (3)$$

$$\ddot{\mathbf{r}}_{sj} = +w^2 \mathbf{r}_{sj} - 2\Gamma\dot{\mathbf{r}}_{sj} - w^2\mathbf{R} \qquad (4)$$

$$b_j \equiv 2m_j\Gamma \qquad w^2 \equiv MJ \qquad (5)$$

The subscripts of "a" (attractive) and "s" (repulsive = spreading) have been introduced to distinguish the two types of systems, "dots" denote derivatives with respect to time, and $\mathbf{V}$ is the velocity of the center of mass. The solution for $\mathbf{R}(t)$ is found by direct integration of Eq. (2)

$$\mathbf{R}(t) = \mathbf{R}_o + \left(\frac{\mathbf{V}_o}{2\Gamma}\right)(1 - e^{-2\Gamma t}) \equiv \mathbf{R}_f - \left(\frac{\mathbf{V}_o}{2\Gamma}\right)e^{-2\Gamma t} \qquad (6)$$

where the subscripts "o" and "f" indicate initial and final values. Changing variables to $\{Z_{aj}\}$ and $\{Z_{sj}\}$ which represent the vector from $r_j$ to $R_f$ gives

$$Z_j \equiv r_j - R_f \qquad \dot{Z} = \dot{r}_j \qquad \ddot{Z} \equiv \ddot{r}_j \qquad (7)$$

$$\ddot{Z}_{aj} = -w^2 Z_{aj} - 2\Gamma \dot{Z}_{aj} - w^2 \left(\frac{V_o}{2\Gamma}\right) e^{-2\Gamma t} \qquad (8)$$

$$\ddot{Z}_{sj} = +w^2 Z_{sj} - 2\Gamma \dot{Z}_{sj} + w^2 \left(\frac{V_o}{2\Gamma}\right) e^{-2\Gamma t} \qquad (9)$$

The solutions for $Z(t)$ are of the form of a "homogeneous" solution (the $V_o = 0$ case) plus a "particular" solution. The particular solutions are found by direct integration of the last terms in Eqs. (8) and (9) respectively

$$Z_{ap} = -\left(\frac{w^2 V_o}{8\Gamma^3}\right)(e^{-2\Gamma t} - 1) - \left(\frac{w^2 V_o}{4\Gamma^2}\right) t \qquad (10)$$

$$Z_{sp} = +\left(\frac{w^2 V_o}{8\Gamma^3}\right)(e^{-2\Gamma t} - 1) + \left(\frac{w^2 V_o}{4\Gamma^2}\right) t \qquad (11)$$

### IIa. Attractive Systems

The expression in Eq. (8) is recognized as a forced, damped simple harmonic oscillator.[2-4] which allows three different qualitative homogeneous solutions: Underdamped, Overdamped, and Critically Damped.[2] As the critically damped case is a special case, only the Overdamped $Z_{aO}$ and Underdamped $Z_{aU}$ solutions will be presented here. For the Underdamped case

$$Z_{aUj} = e^{-\Gamma t}\left(A_j \cos(\Omega t) + B_j \sin(\Omega t)\right) + C_j t + D_j \qquad (12)$$

$$\Omega^2 \equiv w^2 - \Gamma^2 \qquad (13)$$

The constants $C_i$ and $D_i$ allow for an overall choice of any inertial coordinate system. Here $C_i$ is chosen so that the term proportional to t in the full solution is zero. Also $D_j$ is chosen so that each $Z_{aj}(t)$ meets the condition that it approaches zero as t approaches infinity,[5]

$$Z_{aUj} = e^{-\Gamma t}\left(A_j \cos(\Omega t) + B_j \sin(\Omega t)\right) - \left(\frac{w^2 V_o}{8\Gamma^3}\right) e^{-2\Gamma t} \qquad (14)$$

$$r_{aUj} = e^{-\Gamma t}\left(A_j \cos(\Omega t) + B_j \sin(\Omega t)\right) - \left(\frac{w^2 V_o}{8\Gamma^3}\right) e^{-2\Gamma t} + R_o + \left(\frac{V_o}{2\Gamma}\right) \qquad (15)$$

The values of $A_j$ and $B_j$ are determined by the initial conditions of each particle. For the attractive overdamped case, one obtains

$$Z_{aOj} = A_j e^{-(\Gamma - \gamma)t} + B_j e^{-(\Gamma + \gamma)t} - \left(\frac{w^2 V_o}{8\Gamma^3}\right) e^{-2\Gamma t} \qquad (16)$$

$$r_{aOj} = A_j e^{-(\Gamma - \gamma)t} + B_j e^{-(\Gamma + \gamma)t} - \left(\frac{w^2 V_o}{8\Gamma^3}\right) e^{-2\Gamma t} + R_o + \left(\frac{V_o}{2\Gamma}\right) \qquad (17)$$

$$\gamma^2 \equiv \Gamma^2 - w^2 \qquad (18)$$

## IIa. Repulsive Systems

For repulsive systems described by Eq. (9) the homogeneous equation is

$$\ddot{Z}_{sj} - w^2 Z_{sj} + 2\Gamma \dot{Z}_{sj} \equiv 0 \tag{19}$$

Assuming a solution of exponential form and substituting gives

$$Z_{sj}(t) = A_j e^{\lambda t} \tag{20}$$

$$\lambda^2 - w^2 + 2\Gamma\lambda \equiv 0 \tag{21}$$

$$\lambda_+ = (\Gamma^2 + w^2)^{1/2} - \Gamma > 0 \qquad \lambda_- = -\left[(\Gamma^2 + w^2)^{\frac{1}{2}} + \Gamma\right] < 0 \tag{22}$$

The position of each particle is

$$\boldsymbol{r}_{sj} = \left(A_j e^{\lambda_+ t} + B_j e^{-\lambda_+ t}\right) + \left(\frac{w^2 V_o}{8\Gamma^3}\right) e^{-2\Gamma t} - \boldsymbol{R}_o - \left(\frac{V_o}{2\Gamma}\right) \tag{23}$$

and as the time becomes "large" we note that

$$\boldsymbol{r}_{sj}(t \to \infty) = A_j e^{\lambda_+ t} - \boldsymbol{R}_o - \left(\frac{V_o}{2\Gamma}\right) \qquad \boldsymbol{v}_{sj}(t \to \infty) = \lambda_+ A_j e^{\lambda_+ t} \tag{24}$$

Despite an increasingly large wind resistance force every particle moves away from the origin **with an exponentially increasing speed**. For every component, for example the x-component, one also finds that

$$\frac{(v_{sj})_x}{(r_{sj})_x}(t \to \infty) = \lambda_+ \tag{25}$$

which fixes the ratio of potential energy to kinetic energy.

## III. Single Particle Linear Drag

A system with N particles, but with only **one** particle, numbered "n," subject to the linear wind resistance, $b_n = b$, allows Eq. (2) to be integrated directly

$$M \frac{d}{dt} \boldsymbol{V} = -b \frac{d}{dt} \boldsymbol{r}_n \tag{26}$$

$$\boldsymbol{V} - \boldsymbol{V}_o = -\frac{b}{M}(\boldsymbol{r}_n - \boldsymbol{r}_{no}) \tag{27}$$

Because **every** particle interacts directly with **every** other particle, if **any** particle is subject to linear wind resistance, then it is necessarily true that

$$\boldsymbol{V}(t \to \infty) = 0 \tag{28}$$

$$\boldsymbol{r}_n(t \to \infty) = \boldsymbol{r}_{no} + \frac{M\boldsymbol{V}_o}{b} \tag{29}$$

As known previously, in an attractive system with $\boldsymbol{V}_o = 0$ all of the masses will end up at $\boldsymbol{r}_{no}$, regardless of N, the values of the individual masses, or the initial conditions of the other N-1 masses.[5]

The expression in Eqs. (28) and (29) are **also** true for **repulsive** systems. The **one** particle "afflicted" with wind resistance in a repulsive system **always remains "local,"** it is the **only** particle that remains local, and the final position is its initial position plus the displacement of the center of mass. The remaining N-1 particles must move infinitely far away, while the center of mass moves according to Eq. (6).

\* joseph.west@indstate.edu